\documentclass{ws-p8-50x6-00}

\usepackage{amsmath,amssymb}
\usepackage[english]{babel}

\newcommand{\EtAl} {\emph{et~al}}
\newcommand{\GeV}  {{\text{~GeV}}}
\newcommand{\CN}   {{\tilde\chi^{\pm,0}}}

\begin{document}

\title{Bounds on extra quark-lepton generations from precision measurements}

\author{M.~Maltoni}

\address{
  Instituto de F\'{\i}sica Corpuscular~--~C.S.I.C., \\
  Departament de F\'{\i}sica Te\`orica, Universitat de Val\`encia, \\
  Edificio Institutos de Paterna, Apt.~2085, E-46071 Valencia, Spain \\
  INFN, Sezione di Ferrara, Via Paradiso 12, I-44100 Ferrara, Italy \\
  E-mail: maltoni@ific.uv.es}

\maketitle

\abstracts {
  The existence of extra chiral generations is strongly disfavored by the
  precision electroweak data if all the extra fermions are heavier than $m_Z$.
  However fits as good as the SM can be obtained if one allows the new neutral
  leptons to have masses close to $50\GeV$. In the framework of SUSY models
  precision measurements cannot exclude one additional generation of heavy
  fermions if chargino and neutralino have masses around $60\GeV$ with $\Delta
  m_\CN \simeq 1\GeV$.}

\section{Introduction}

The aim of this talk is to analyze to what extent precision measurements of
IVB parameters allow to bound effectively the existence of extra chiral
generations of heavy fermions, both quarks ($q=U,D$) and leptons
($l=N,E$).~\cite{Maltoni00} We will show that the case where \emph{all} the
extra particles are heavier than $m_Z$ is now excluded by more than $2$
standard deviations. However, if the masses of new neutrinos are assumed to be
close to $50\GeV$, then additional generations become allowed and up to three
extra families can exist within $2.5\sigma$. Finally, inclusion of new
generations in SUSY extension of Standard Model is briefly discussed.

For simplicity, we will assume that the extra neutrinos are just ordinary
massive Dirac particles, and that their mass is larger than $45\GeV$ so to
avoid contradiction with the experimental data on the invisible $Z$ width.
Also, we will neglect the possible mixing among new generations and the three
existing ones, hence new fermions contribute to electroweak observables only
through oblique corrections. On the other hand, we perform global fit of
\emph{all} precision data,~\cite{NORV99} studying both degenerate and
non-degenerate extra generations on the equal footing. Taking the number of new
generations $N_g$ as a continuous parameter, just as it was done with the
determination of the number of neutrinos from invisible $Z$ width, we get a
bound on it. The minimum $\chi^2$ corresponds to $N_g\simeq -0.5$, while the
case $N_g=1$ is excluded by more than $2$ standard deviations.

\section{Discussion}

\begin{figure} \centering
    \begin{minipage}[t]{0.485\textwidth}
	\epsfxsize=\textwidth \epsfbox{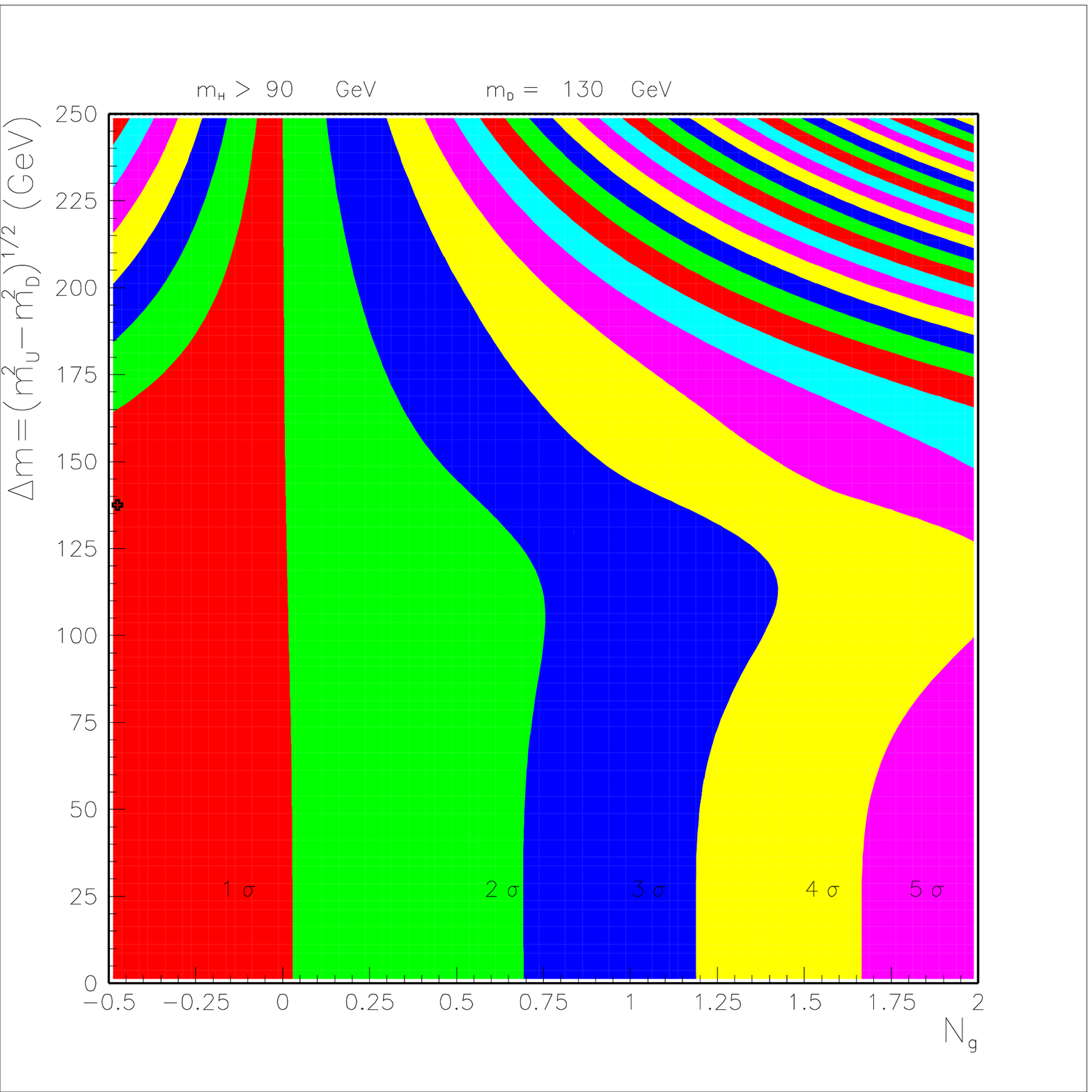}
	\caption{\label{fig:20} 
	  Exclusion plot in the $(N_g, \Delta m)$ plane. We assumed $m_N =
	  m_U$ and $m_E = m_D = 130\GeV$ (Tevatron bound). Regions show
	  $<1\sigma$, $<2\sigma$ etc.\ allowed domains.}
    \end{minipage} \hfill
    \begin{minipage}[t]{0.485\textwidth}
	\epsfxsize=\textwidth \epsfbox{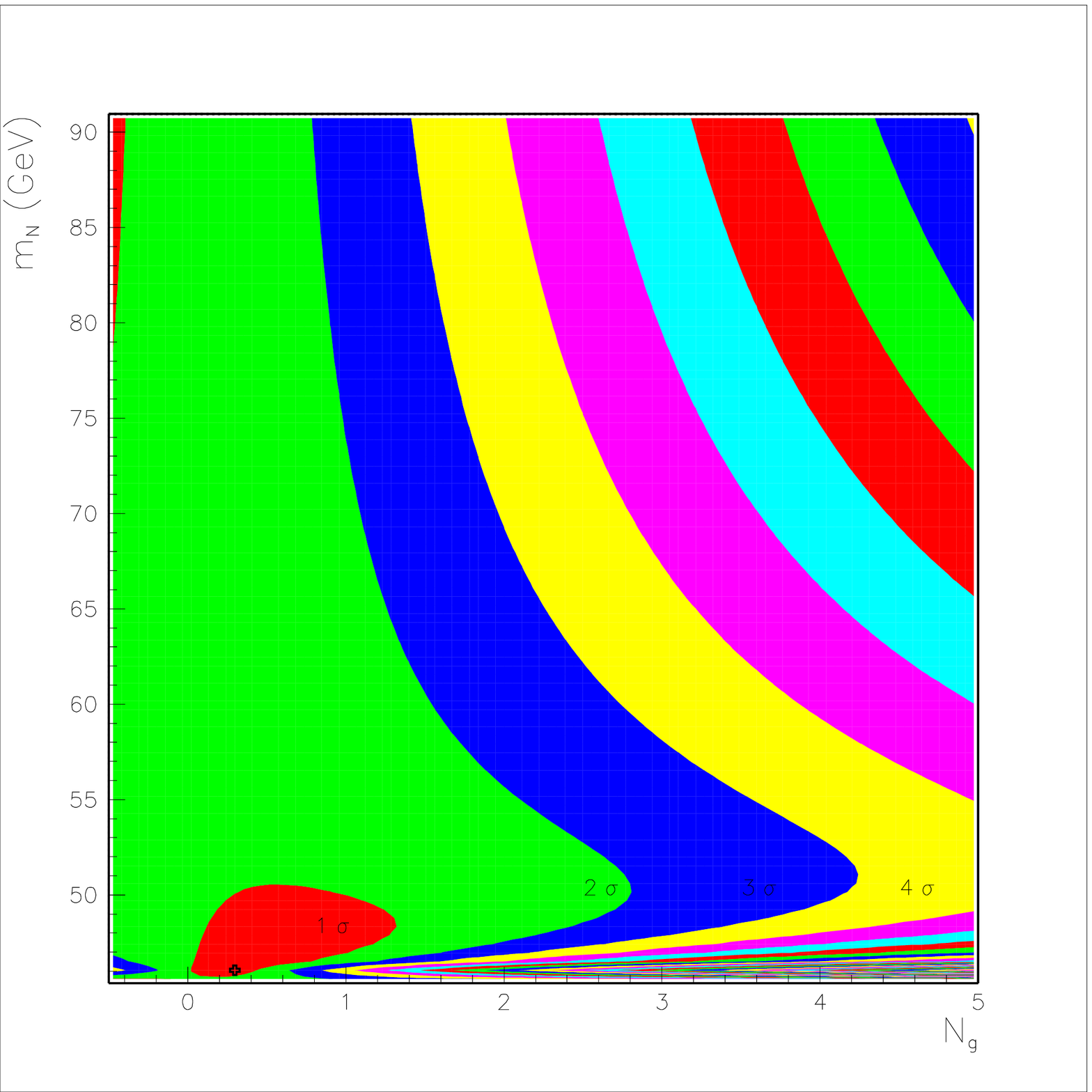}
	\caption{\label{fig:30}
	  Exclusion plot in the $(N_g, m_N)$ plane. We assumed $m_U =
	  220\GeV$, $m_D = 200\GeV$, $m_E = 100\GeV$. Regions show $<1\sigma$,
	  $<2\sigma$ etc.\ allowed domains.}
    \end{minipage}
\end{figure}

\noindent\textit{Heavy fermions.}
The comparison between theoretical predictions and experimental data is
performed with the help of the computer code
\texttt{LEPTOP}.~\cite{NOV95,NORV99} To simplify the analysis we start from
the ``horizontally degenerate'' case $(m_N = m_U) > (m_E = m_D =
130\GeV)$,~\cite{Maltoni00,MaltoniPHD} the lower bound on extra quark masses
coming from Tevatron search,~\cite{PDG98} and in Fig.~\ref{fig:20} we show the
excluded domains in coordinates ($N_g$, $\Delta m \equiv (m_U^2 -
m_D^2)^{1/2}$). Any value of higgs mass above $90\GeV$ is allowed in our fits;
minimum of $\chi^2$ corresponds to $N_g =-0.5$, and the case $N_g = 0$ is
within the $1\sigma$ domain. It is immediate to see that one extra generation
is always excluded by more than $2.5$ standard deviations.

We checked that similar bounds are valid for the general choice of heavy
masses of leptons and quarks.~\cite{MaltoniPHD} In particular we found that
for the ``cross degenerate'' case $(m_E = m_U) > (m_N = m_D = 130\GeV$) one
extra generation is excluded at $2\sigma$ level, while in the ``anti-cross
degenerate'' case $(m_N = m_D) > (m_E = m_U = 130\GeV)$ the limits are even
stronger than in Fig.~\ref{fig:20}.

\medskip\noindent\textit{Light neutrinos.}
For particles with masses close to $m_Z/2$, oblique corrections drastically
differ from what we have above $m_Z$. If we assume that the mixing angle
between $N$ and the three known neutrinos is less than $10^{-6}$, so to avoid
the bound $m_N > 70\div 80\GeV$ from LEP~II searches of the decays $N\to
lW^{\ast}$, we have that extra neutrinos as light as $45\GeV$ are still
allowed by direct search experiments. In this case, effects of $Z$-boson wave
function renormalization become relevant,~\cite{Evans94} and the quality of
the fit can be even better than the SM. Analyzing all the electroweak
observables, we conclude that presently a light neutral lepton $N$ cannot be
excluded by precision measurements as well. As an example, in
Fig.~\ref{fig:30} we assume $m_U = 220\GeV$, $m_D = 200\GeV$, $m_E = 100\GeV$
and draw the exclusion plot in coordinates ($m_N, N_g$): from this plot it is
clear that for the case of extra generations with $m_N\approx 50\GeV$ even two
new generations are allowed within $1.5\sigma$.

\begin{figure} \centering
    \begin{minipage}[t]{0.485\textwidth}
	\epsfxsize=\textwidth \epsfbox{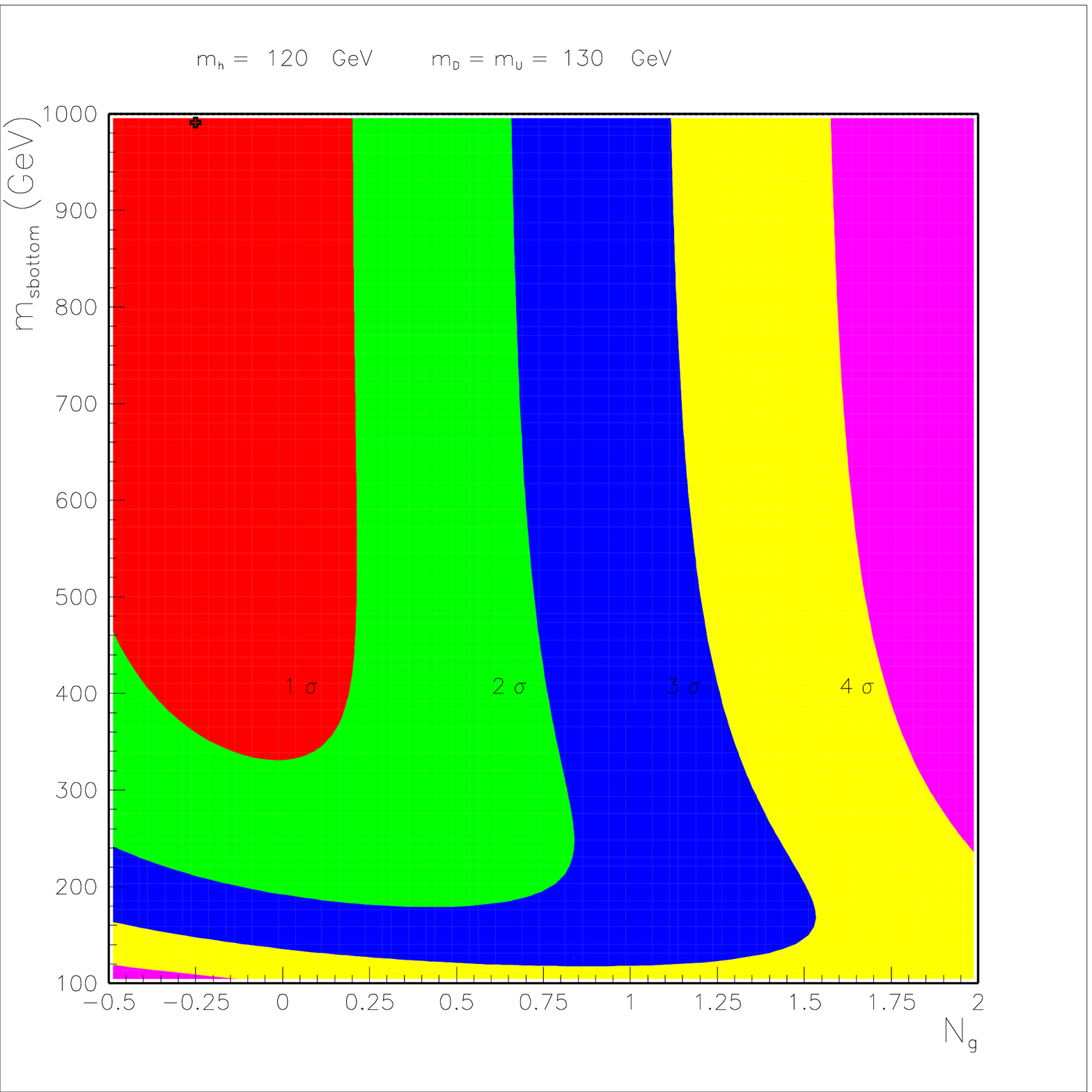}
	\caption{\label{fig:40}
	  Exclusion plot in the $(N_g, m_{\tilde{b}})$ plane. We assumed
	  $m_{N,E,U,D} = 130\GeV$, $m_{\tilde{g}} = 200\GeV$ and no $\tilde
	  {t}_L - \tilde {t}_R$ mixing. Regions show $<1\sigma$, $<2\sigma$
	  etc.\ allowed domains.}
    \end{minipage} \hfill
    \begin{minipage}[t]{0.485\textwidth}
	\epsfxsize=\textwidth \epsfbox{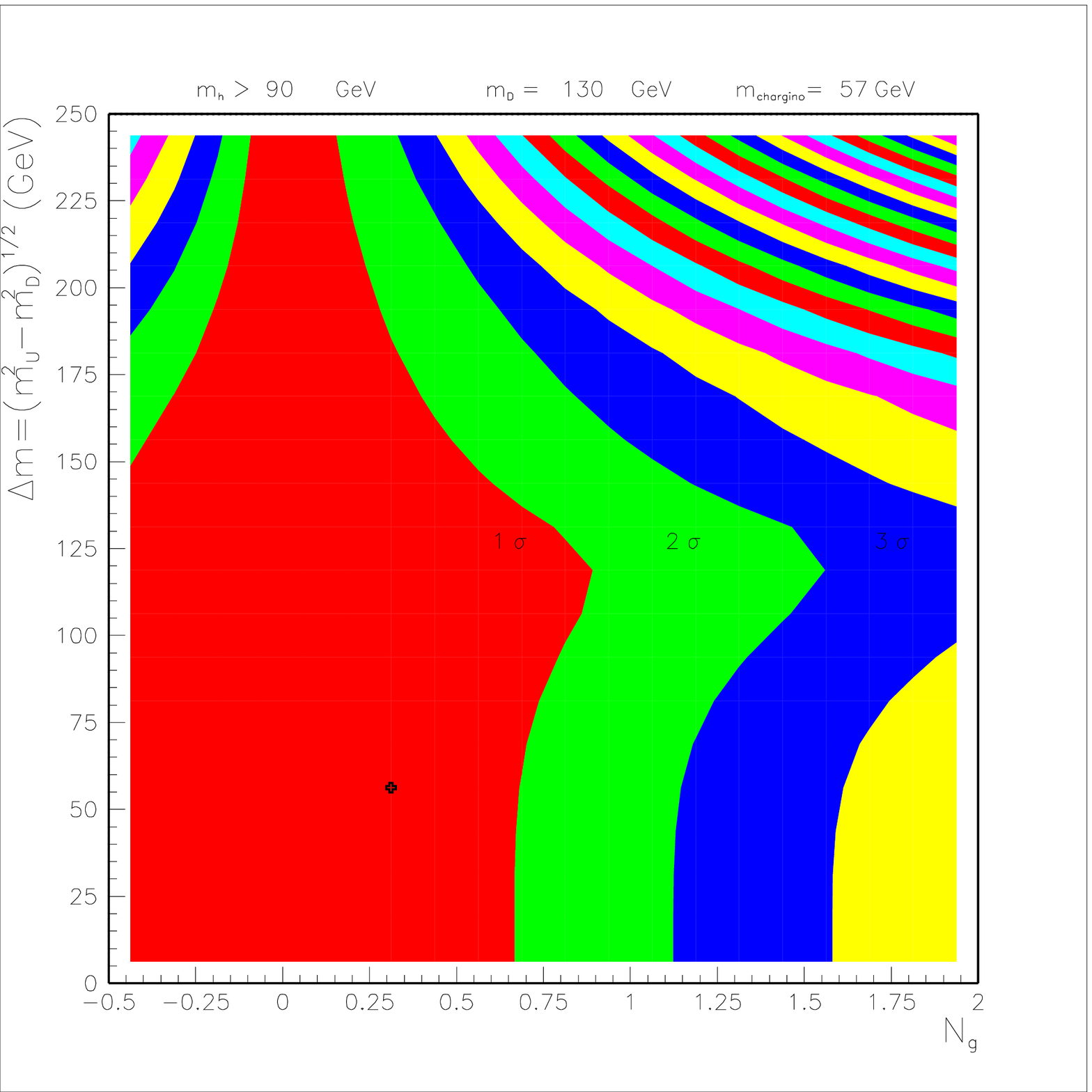}
	\caption{\label{fig:50}
	  Same as Fig.~\ref{fig:20}, but including the extra contributions
	  from a $57\GeV$ almost degenerate chargino and neutralino in the
	  higgsino-dominated scenario.}
    \end{minipage}
\end{figure}

\medskip\noindent\textit{The case of SUSY.}
Concerning bounds on extra generations which occur in SUSY extensions, when
superpartners are heavy their contributions to electroweak observables become
power suppressed, and the same Standard Model exclusion plots shown in
Fig.~\ref{fig:20} and Fig.~\ref{fig:30} are valid. The present lower bounds on
the sparticle masses from direct searches leave mainly this decoupled domain.
One possible exception is a contribution of the third generation squark
doublet, enhanced by large stop-sbottom splitting.~\cite{Gaidaenko98} To see
whether this contribution affects the bounds on extra generations found in the
previous section, in Fig.~\ref{fig:40} we analyze the simplest case of the
absence of $\tilde {t}_L - \tilde {t}_R $ mixing, setting the masses of all
the extra fermions to the common value $m_{N,E,U,D} = 130\GeV$ and showing the
plot in coordinates $(N_g, m_{\tilde{b}})$. It is clear that even in the
context of SUSY models new heavy generations are disfavored.

Situation qualitatively changes in the case of almost degenerate light
chargino and neutralino. This possibility is yet not excluded~-- dedicated
search at LEP~II by DELPHI still allows the existence of such particles with
masses as low as $45\GeV$ if their mass difference is $\approx
1\GeV$~\cite{DELPHI99}~-- and even in this case radiative corrections are
large.~\cite{Maltoni99} Fig.~\ref{fig:50} demonstrates how the presence of a
chargino-neutralino pair (dominated by higgsino) with mass $57\GeV$ relaxes
the bounds shown on Fig.~\ref{fig:20}: we see that one extra generation of
heavy fermions is now allowed within the $1.5\sigma$ domain.

\section{Conclusions}

Inclusion of new generations in Standard Model is not excluded by precision
data if the new neutral leptons have masses close to $m_Z/2$ (see
Fig.~\ref{fig:30}). In order to experimentally investigate this case a special
search for the reaction $e^+ e^- \rightarrow \gamma Z^* \rightarrow \gamma
N\bar{N}$ with larger statistics and improved systematics is needed. Finally,
further experimental search for light chargino and neutralino~\cite{DELPHI99}
is of interest. These searches could close the existing windows of ``light''
extra particles, or open a door into a realm of New Physics.

\section*{Acknowledgments}

I wish to thank my collaborators V.A.~Novikov, A.N.~Rozanov, L.B.~Okun and
M.I.~Vysotsky. This work was supported by DGICYT under grant PB98-0693 and by
the TMR network grant ERBFMRX-CT96-0090 of the European Union.


\begin{thebibliography}{99}

\bibitem{Maltoni00}
  M.~Maltoni \EtAl, \Journal{\PLB}{476}{107}{2000}, and references therein.

\bibitem{NORV99}
  V.A.~Novikov \EtAl, \Journal{\em Rept.\ Prog.\ Phys.}{62}{1275}{1999}.

\bibitem{NOV95}
  V.A.~Novikov \EtAl, preprints {\em ITEP-19-95}, {\em CPPM-1-95}.

\bibitem{MaltoniPHD}
  M.~Maltoni, Ph.~D.\ thesis, {\em hep-ph/0002143}, Ch.~5 (2000).

\bibitem{PDG98} Review of Particle Physics, 
  {\em Eur.\ Phys.\ J.} C {\bf 3}, No.~1-4 (1998).

\bibitem{Evans94}
  N.~Evans, \Journal{\PLB}{340}{81}{1994}; \\
  P.~Bamert, C.P.~Burgess, \Journal{\ZPC}{66}{495}{1995}.

\bibitem{Gaidaenko98}
  I.V.~Gaidaenko \EtAl, \Journal{\em JETP Lett.}{67}{761}{1998};
  \Journal{\em Phys.\ Rept.}{320}{119}{1999}.

\bibitem{DELPHI99}
  DELPHI Coll., P.~Abreu \EtAl, \Journal{{\em Eur.\ Phys.\ J.} C}{11}{1}{1999}.

\bibitem{Maltoni99} 
  M.~Maltoni, M.I.~Vysotsky, \Journal{\PLB}{463}{230}{1999}.

\end{thebibliography}
\end{document}